
\catcode`\@=11
\def\oldcases#1{\left\{\,\vcenter{\normalbaselines\openup\jot\m@th
    \ialign{$##\hfil$&\quad##\hfil\crcr#1\crcr}}\right.}
\catcode`\@=12 

\input amstex
\documentstyle{amsppt}
\nologo
\pagewidth{160truemm}
\pageheight{240truemm}

\def\tg{{\tilde g}}
\def\tR{{\tilde R}}
\def\dirac#1{\delta^{(#1)}}
\def\Real{{\Bbb R}}
\def\Nat{{\Bbb N}}
\def\Zint{{\Bbb Z}}
\let\eps=\varepsilon
\let\leq=\leqslant

\let\geq=\geqslant

\let\cal=\Cal
\let\embed=\iota

\def\cD{{\cal D}}               
\def\CA#1{{\cal A}_{#1}}        
\def\CE{{\cal E}}               
\def\CM{{\cal E}_M}             
\def\CN{{\cal N}}               
\def\CG{{\cal G}}               
\def\norm#1{\left|#1\right|}
\def\restrict#1#2{\left.#1\right|_{#2}}
\def\supp{\operatorname{supp}}
\def\pbyp#1#2{{\partial#1\over\partial#2}}
\def\x{\lambda}
\def\y{\mu}
\def\dbyd#1#2{{d#1 \over d#2}}
\def\emph#1{{\it#1}}

\topmatter
\title
Invariance of the distributional curvature of the cone under smooth
diffeomorphisms
\endtitle
\author
J.A.Vickers and J.P.Wilson
\endauthor
\address
Faculty of Mathematical Studies,
University of Southampton,
Southampton  SO17 1BJ, UK.
\endaddress
\email\nofrills
\emph{Email addresses}:
jav\@maths.soton.ac.uk,
jpw\@maths.soton.ac.uk
\endemail
\subjclass\nofrills
PACS: 0420, 1127
\endsubjclass
\abstract
An explicit calculation is carried out to show that the distributional
curvature of a 2-cone, calculated by Clarke et al.\ (1996), using
Colombeau's new generalised functions is invariant under non-linear
$C^\infty$ coordinate transformations.
\endabstract
\rightheadtext{Invariance of the distributional curvature of the cone}
\endtopmatter

\document  

\head
1. Introduction
\endhead
Recently, Colombeau's theory of new generalised functions (Colombeau, 1984)
has been applied to give a distributional interpretation to physical
quantities in General Relativity that are calculated by non-linear
processes, such as the distributional curvature of cosmic strings (Clarke
et al, 1996), the Ultra-relativistic Riessner Nordstr{\o}m field
(Steinbauer, 1997) and the energy-momentum tensor of the Kerr solution
(Balasin, 1997).  One of the major problems of using the full Colombeau
algebra, which undermines its usefulness in a covariant physical theory
such as General Relativity is that it is not of a coordinate invariant
construction; that is given a diffeomorphism $\mu:\Omega\to\Omega'$ between
two open sets, there is no natural way of defining a map
$\tilde\mu^*:\CE(\Omega')\to\CE(\Omega)$ which preserves the concepts of
moderate and null functions, and commutes with the canonical embedding
(smoothing convolution) $\embed:\cD'(\Omega)\to\CE(\Omega)$.

In this paper we shall return to the calculations of Clarke et al.\ (1996)
and we shall explicitly show that the distributional curvature is
independent of any non-linear $C^\infty$ coordinate transformation that we
might carry out on the metric. This will provide evidence for a covariant
formalism of Generalised functions

\head
2. The distributional curvature of cosmic strings
\endhead
The distributional curvature of the cone with a deficit angle of
$2\pi(1-A)$, whose metric may be written as
$$ \eqalign{
   & g_{ab} = \tfrac12 (1+A^2) \delta_{ab} + \tfrac12 (1-A^2) m_{ab} \cr 
   & m_{ab} = \pmatrix {x{}^2-y{}^2\over x{}^2+y{}^2} & {2xy\over
   x{}^2+y{}^2} \cr 
   {2xy\over x{}^2+y{}^2} & -{x{}^2-y{}^2\over x{}^2+y{}^2}
   \cr\endpmatrix  \cr
}\eqno(1)$$
was recently calculated by Clarke et al.\ (1996) by using Colombeau's theory
of generalised functions to overcome the problem of assigning
distributional interpretations to products of distributions. It was shown
to be
$$ R\sqrt{g}= 4\pi(1-A)\dirac2(x,y) $$
This result implies that a thin cosmic string, whose exterior metric is
$$ ds^2=-dt^2+dr^2+A^2r^2d\phi^2+dz^2 $$
has a mass per unit length of $2\pi(1-A)/A$.

The process used to calculate the curvature first involves an embedding of
the metric $g_{ab}$ into $\CE_M(\Real^2)$;
$$ \eqalign{
   & \embed :g_{ab} \mapsto \widetilde{g_{ab}} \cr
   & \widetilde{g_{ab}}(\Phi,x,y) = \int g_{ab}(x+\eps\xi,y+\eps\eta)
   \Phi(\xi,\eta) \,d\xi\,d\eta \cr
}$$
The generalised function Ricci curvature $\tR\sqrt{\tg}$ is then is
calculated from the metric $\widetilde{g_{ab}}$ in the usual manner, and
finally it is shown that this curvature is equivalent to the distribution
$4\pi(1-A)\dirac2$ in the sense of weak equivalence; that is that for each
$\Psi\in\cD(\Real^2)$
$$ \lim_{\eps\to0} \int \tR\sqrt\tg \Psi \,dx\,dy = \Psi(0,0) $$  
for $\Phi\in\CA{q}(\Real^2)$, for large enough $q\in\Nat$.

The theory of General Relativity is however a covariant theory, so we would
like these results to make sense in the context of coordinate invariance;
that is given a $C^\infty$ diffeomorphism;
$$ \eqalign{ & \mu:\Real^2\to\Real^2 \cr & (x',y')=\mu(x,y) \cr } $$
the following diagram commutes;
$$
\CD
   {g'_{ab}} @>{\embed'}>> {\widetilde{g'_{ab}}} @>>>
   {\tR'\sqrt{\tg'}} @>{\approx}>> {4\pi(1-A)\delta^{(2)}} \\
   @VV{\mu^*}V &&&&  @VV{\mu^*}V \\
   {g_{ab}} @>{\embed}>> {\widetilde{g_{ab}}} @>>>
   {\tR\sqrt{\tg}} @>{\approx}>> {4\pi(1-A)\delta^{(2)}}
\endCD
$$
where $\mu^*$ denotes the appropriate tensor or density transformation laws

Since the full Colombeau algebra is not a coordinate invariant construction
and the embedding $\embed$ does not commute with $\mu$, it will only make sense
to talk about coordinate invariance at the level of distribution theory;
this will mean that we must follow the approach of carrying out the
coordinate transformation on the metric $g'_{ab}$ to form a new metric
$g_{ab}$, explicitly calculate the distributional Ricci scalar density for
the new metric and show that it is transforms as a scalar density of weight
$+1$ from the Ricci scalar density that corresponded to $g'_{ab}$.

We first apply $C^\infty$ diffeomorphism $\mu:\Real^2\to\Real^2$ to the
primed version of the metric~(1); The fact that Colombeau's theory is
manifestly invariant under linear coordinate transformations means that
without loss of generality, it may be assumed that $\mu$ is of the form
$$ \eqalign{ x' &= x+f(x,y) \cr y' &= y+g(x,y) \cr } \eqno(2)$$
where $f$ and $g$ are $C^\infty$ and $O(x^2+y^2)$.

On applying~(2) to~(1) we are able to write
$$ g_{ab}(x,y) = \tfrac12(1+A^2) l_{ab}(x,y)  +
   \tfrac12(1-A^2)  m_{ab}(x,y) $$
where
$$ \eqalign{
   l_{ab} &= \pbyp{x'{}^c}{x^a} \pbyp{x'{}^d}{x^b} \delta_{cd} \cr
   m_{ab} &= \pbyp{x'{}^c}{x^a} \pbyp{x'{}^d}{x^b} m'_{cd} \cr}$$
and
$$ \def\tempa#1{{#1\over (x+f)^2+(y+g)^2}}
    \eqalign{
   m'_{ab} &= \pmatrix \tempa{(x+f)^2-(y+g)^2} &
   \tempa{2(x+f)(y+g)} \cr \tempa{2(x+f)(y+g)} &
   -\tempa{(x+f)^2-(y+g)^2} \cr \endpmatrix \cr
   \pbyp{x'{}^a}{x^b} &= \pmatrix 1+f_{,x} & f_{,y} \cr g_{,y} & 1+g_{,y}
   \cr \endpmatrix \cr
}$$
Next we shall smooth $g_{ab}$ with a kernel $\Phi\in\CA0(\Real^2)$ having a
support in a radius of
$$ R_0 = \bigl\{\,(x^2+y^2)^{1/2} \,\bigm|\,|\Phi(x,y)|>0\,\bigr\}. $$
The functions $l_{ab}$, consisting of products of the smooth Jacobian
components, will be $C^\infty$ and so may be identified with their
smoothings in $\CM(\Real^2)$.

The matrix $m_{ab}$, containing the singular contributions to the metric,
may be expressed as
$$ \def\tempa#1{{#1\over x^2+y^2}}
   m_{ab} = \left( 1+2\tempa{xf+yg}+\tempa{f^2+g^2} \right)^{-1}
   \pmatrix
   \tempa{x^2-y^2+f_{11}} & \tempa{2xy + f_{12}} \cr
   \tempa{2xy + f_{21}} & \tempa{y^2-x^2+f_{22}} \cr
   \endpmatrix
$$
where the $f_{ab}(x,y)$ are $C^\infty$ and $O(r^3)$ functions.
It may be observed that the components of the above matrix may be
expressed as linear combinations of real and imaginary parts of
complex-valued functions having the form 
$$ m(x,y) = {e^{2i\phi} + m_1(x,y) e^{-2i\phi} + m_2(x,y) +
   m_3(x,y) e^{2i\phi} \over 1 + m_4(x,y) e^{-2i\phi} +
   m_5(x,y) + m_6(x,y) e^{2i\phi}} $$
where we define the polar coordinates $(r,\phi)$ by
$$ x=r\cos\phi, \quad y=r\sin\phi; $$
and the $m_k(x,y)$ are $C^\infty$ and $O(r)$.  Therefore we may
calculate $\widetilde{m_{ab}}_\eps$ by smoothing $h(x,y)$.

The expression for  $m(x,y)$ may be expanded as
$$ m(x,y)= e^{2i\phi} + r \!\!\! \sum_{k=\pm1,\pm3} \!\!\! \alpha_k
   e^{ki\phi} + r^2 p(x,y) \eqno(3)$$
where the $\alpha_k$ are constants and $p(x,y)$ is $O(r^0)$, continuous
everywhere and $C^\infty$ everywhere except at $r=0$.  We may smooth this
expansion termwise and obtain estimates for the smoothings. Certainly the
$re^{\pm i\phi}$ are $C^\infty$ and may be identified with their
smoothings. For other terms, the results in the appendix will give
$$ \eqalign{
   \widetilde{e^{2i\phi}} &= \oldcases{
   C_1 + C_2 x/\eps + C_3 y/\eps + O(r^2/\eps^2) & if $r<\eps R_0$ \cr
   e^{2i\phi} \left( 1+ O\biggl({\eps^{q+1} \over r^{q+1}}\biggr)
   \right) & if $r>\eps R_0$ \cr
}\cr
   \widetilde{re^{\pm3i\phi}} &= \oldcases{
   \eps\bigl( C_{\pm4} + C_{\pm5} x/\eps + C_{\pm6} y/\eps +
   O(r^2/\eps^2) \bigr) & if $r<\eps R_0$ \cr
   re^{\pm3i\phi} \left( 1+ O\biggl({\eps^{q+1} \over r^{q+1}} \biggr)
   \right) & if $r>\eps R_0$ \cr
}\cr
}$$ 
where the $C_k$ are constants.

The only smoothing that we have not yet had to estimate is that of the
remainder term $r^2 p(x,y)$. This term will provide a higher order
contribution to the metric and therefore will not require a such a delicate
estimate in order to be able to differentiate it when we come to calculate
the curvature. We may write the smoothing of this term as
$$ \eqalign{%
   \widetilde{r^2 p(x,y)} &= \int^\infty_0 \int^{2\pi}_0
   \bigl(r^2+2\eps\rho\cos(\psi-\phi)+\eps^2\rho^2\bigr) \cr
   &\qquad\times p(r\cos\phi +\eps\rho\cos\psi, r\sin\phi +
   \eps\rho\sin\psi) \cr
   &\qquad\times\Phi(\rho\cos\psi,\rho\sin\psi) \rho\,d\rho\,d\psi
} $$

For $r<\eps R_0$, we have
$$ \eqalign{
   \restrict{\widetilde{r^2p(x,y)}}{r=0} &= \eps^2 \int_{\Real^2}
   (\x^2+\y^2) p(\eps\x,\eps\y) \Phi(\x,\y) \,d\x\,d\y = O(\eps^2),
   \cr
   \restrict{\dbyd{}{x}\left\{\widetilde{r^2p(x,y)}\right\}}{r=0} &=
   2\eps\int_{\Real^2} \x p(\eps\x,\eps\y) \Phi(\x,\y) \,d\x\,d\y \cr
   &\quad + \eps^2 \int_{\Real^2} (\x^2+\y^2) p_{,x}(\eps\x,\eps\y)
   \Phi(\x,\y) \,d\x\,d\y = O(\eps), \cr
}$$
and similarly
$$ \restrict{\dbyd{}{y}\left\{\widetilde{r^2p(x,y)}\right\}}{r=0} =
   O(\eps). $$ 
Therefore
$$ \widetilde{r^2p(x,y)} = M(\eps) + O(\eps r) $$
where $M(\eps)$ is an $O(\eps^2)$ function of $\eps$ alone.

If however, $r>\eps R_0$, then
$$ \eqalign{%
   \widetilde{r^2 p(x,y)} &= r^2\int^\infty_0 \int^{2\pi}_0
   \left(1+2{\eps\rho\over r}\cos(\psi-\phi)+{\eps^2\rho^2\over
   r^2}\right) \cr
   & \qquad\times p\bigl(r(\cos\phi+{\eps\rho/r}\, \cos\psi),
   r(\sin\phi+{\eps\rho/r}\,\sin\psi)\bigr) \cr
   &\qquad\times\Phi(\rho\cos\psi,\rho\sin\psi) \rho\,d\rho\,d\psi%
} $$
and so by  applying the Mean Value Theorem to $f(s)=p(x+s\x,y+s\y)$
$$ p(x+\x,y+\y)=p(x,y)+\x p_{,x}(x+\xi\x,y+\xi\y) + \y
   p_{,y}(x+\xi\x,y+\xi\y), $$
where $\xi\in(0,1)$; it may be concluded that
$$ \widetilde{r^2 p(x,y)} = r^2p(x,y) \bigl(1+O(\eps/r)\bigr). $$

We now may add up the smoothings in the summation~(3) to obtain
$$ \widetilde{h}(x,y) = \oldcases{
   C_1(\eps) + C_2(\eps) {x\over \eps} + C_3(\eps) {y\over\eps} +
   O(r^2/\eps^2) + O(r\eps), & if $r<\eps R_0$, \cr
   h(x,y) \left( 1 + O\biggl({\eps^{q+1}\over r^{q+1}}\biggr)
   \right) + O(r\eps), & if $r>\eps R_0$, \cr
}$$
express the smoothed metric $\widetilde{g_{ab}}$ as
$$ \widetilde{g_{ab}}_\eps =\oldcases{
    C^{(0)}_{ab}(\eps) + C^{(1)}_{ab}(\eps) x/\eps + C^{(2)}_{ab}(\eps)
    y/\eps + O(r^2/\eps^2) + O(r\eps) & if $r<\eps R_0$ \cr
    g_{ab} + O\biggl({\eps^{q+1}\over r^{q+1}}\biggr) + O(r\eps) & if
    $r>\eps R_0$ \cr
}$$
where $C^{(i)}_{ab}(\eps)$ are $O(1)$ functions of $\eps$ alone,
and use it to calculate the Ricci scalar density as
$$ \tR_\eps \sqrt{\tg_\eps} = \oldcases{
   O(1/\eps^2) + O(\eps/r) & if $r<\eps R_0$ \cr
   O(\eps/r^3) & if $r>\eps R_0$ \cr
}$$

To obtain a distributional interpretation to this curvature, we shall
follow the method of Clarke et al.\ (1996). Much of the analysis is
applicable, but this time we shall be working with more crude estimates.

Given $\Psi\in\cD(\Real^2)$, we define
$$ \eqalign{
   K &= \supp\Psi\cr
   R_K &= \sup\bigl\{\,(x^2+y^2)^{1/2}\,\bigm|\,|\Psi(x,y)|>0\,\bigr\} \cr
   B_\eps &= \sup\bigl\{\, (x,y)\in \Real^2 \,\bigm|\, (x^2+y^2)^{1/2}<\eps
   R_0\,\bigr\} \cr 
}$$
By the mean value theorem, we may write for some $\xi\in[0,1]$, that
$$ \int_K \tR_\eps\sqrt{\tg_\eps}(x,y) \Psi(x,y) \,dx\,dy = I_1 +
   I_2 $$
where
$$ \eqalign{
   I_1 &= \int_K \! \tR_\eps\sqrt{\tg_\eps} \,dx\,dy \Psi(0,0) \cr
   I_2 &= \int_K \! \tR_\eps\sqrt{\tg_\eps} \, r \restrict{{d\Psi\over dr}}
   {(\xi x,\xi y)} \!\!\!\! dx\,dy. \cr 
}$$
Now
$$ \eqalign{ \norm{I_2}
    & \leq M_1\int_{B_\eps} \norm{\tR_\eps\sqrt{\tg_\eps}} r\,dx\,dy +
    M_1\int_{K- B_\eps} \norm{\tR_\eps\sqrt{\tg_\eps}} r\,dx\,dy \cr
    & \leq 2\pi M_2 {(\eps R_0)^3 \over \eps^2} + 2\pi M_3 \eps^3 {R_0}^2 +
    2\pi M_4 \eps \bigl( \log{R_k} - \log(\eps R_0)\bigr)
    \cr
}$$
where $M_n$ are positive constants, giving that $I_2=O(\eps\log\eps)$.

To calculate $I_1$ we let $D=\bigl\{\,(x,y)\,\bigm|\,(x^2+y^2)^{1/2}\leq
R_D\,\bigr\}\subseteq K$. and write
$$ I_1 = \int_{K-D} \!\tR_\eps\sqrt{\tg_\eps} \,dx\,dy \, \Psi(0,0)
   +\int_D \! \tR_\eps\sqrt{\tg_\eps} \,dx\,dy \, \Psi(0,0) $$
The first integral will be $O(\eps/{R_D})$, converging to zero as
$\eps\to0$. For the second integral the Gauss-Bonnet theorem is
applicable. It is also the case that
$$ \int_{\partial D} \kappa_{{\tg_\eps}} \,ds = 2\pi A+ O(\eps/R_D)$$
Therefore
$$ I_1= 4\pi(1-A) \Psi(0,0) + O(\eps/R_D). $$
It may therefore be concluded that
$$ \lim_{\eps\to0} \int_K \tR_\eps\sqrt{\tg_\eps}(x,y) \Psi(x,y)
   \,dx\,dy = 4\pi(1-A)\Psi(0,0) $$
giving us
$$ [\tR\sqrt{-\tg}] \approx 4\pi(1-A) \dirac2(x,y). $$

This shows that the distributional curvature associated to the curvature of
the generalised metric transforms as a scalar density of weight $+1$.

\head
3. Conclusion
\endhead
We have now shown that the calculations of Clarke et al. (1996) are
independent under $C^\infty$ diffeomorphisms, which suggests that
Colombeau's theory could be reformulated in a covariant manner.

There have been many attempts at amending Colombeau's theory in order to
achieve coordinate invariance in this sense. One such approach has been to
use the simplified algebra (Biagioni, 1990; Colombeau, 1992) in which one
defines the space base $\CE(\Omega)$ as consisting of functionals dependent
on the regularisation parameter $\eps\in(0,1]$, rather than the smoothing
kernel $\Phi\in\CA0(\Omega)$, thus avoiding the problem of using the spaces
$\CA{q}\Real^n)$ in the definitions of $\CE_M(\Omega)$ and $\CN(\Omega)$;
however this still does not make the embedding, which involves $\Phi$,
commute with $\mu$.

A completely different approach was pursued by Colombeau and Meril~(1994)
in which they substantially redefined the kernel spaces $\CA{q}(\Real^n)$
with weaker moment conditions imposed which does result in the ability to
construct $\CG(\Omega)$ in a coordinate invariant manner together with the
embedding commuting with $\mu$. This formalism enables scalar generalised
functions to be constructed, whose transformation laws coincide with those
of scalar distributions.

Work is currently in progress (Vickers and Wilson, 1998) into extending the
formalism of Colombeau and Meril to enable generalised functions to be
constructed as multi-index tensors whose transformation laws coincide with
those of tensor distributions.

\head
Appendix. Smoothing of polar functions
\endhead
In this appendix we shall smooth the non-regular complex functions
$e^{2i\phi}$ and $re^{3i\phi}$ and hence obtain estimates for the regions
$r<\eps R_0$ and $r>\eps R_0$. For the fine details of calculation the
reader is referred to Clarke et al.\ (1996). We begin with the function
$$ f(x,y) = r^{a-1} e^{i(a+1)\phi} $$
where $a=1,2$.
Its smoothing, with a kernel $\Phi\in\CA{q}(\Real^2)$ may be expressed in
polar coordinates as
$$ {\tilde f}(\Phi,r\cos\phi,r\sin\phi) = \int {(re^{i\phi}+\eps\rho
   e^{i\psi})^a \Phi(\rho\cos\psi,\rho\sin\psi) \over r^2 + \eps^2\rho^2 +
   2\eps\rho \cos(\psi-\phi) } \rho\,d\rho $$
We may expand $\Phi$ as a sum of circular harmonics;
$$ \Phi(r\cos\phi,r\sin\phi) =  \sum_{n\in\Zint} e^{in\phi} $$
so that we may write
$$ {\tilde f}(\Phi,r\cos\phi,r\sin\phi = \sum_{n\in\Zint} F_n(r)
   e^{i(n+a+1)\phi} $$
where
$$ F_n(r) = r^{a-1} \int { 1+\eps\rho/r \, e^{i\psi} \over 1+\eps\rho/r \,
   e^{-i \psi}} e^{in\psi} \Phi_n(\rho) \rho\,d\rho\,d\psi. $$ 
On integrating out $\psi$ we obtain
$$ F_n(r) = \oldcases{
   2\pi r^{a-1} \int^{r/\eps}_0 \left(-{\eps\rho\over r}\right)^n
   \left( 1-{\eps^2\rho^2\over r^2} \right)^a \Phi_n(\rho) \rho\,d\rho & if
   $n\geq0$ \cr
   2\pi r^{a-1} \sum_{k=-n}^a {a\choose k} \int^{r/\eps}_0
   \left(-{\eps\rho\over r}\right)^n \left(-{\eps^2\rho^2\over
   r^2}\right)^k \Phi_n(\rho) \rho\, d\rho \cr
   \quad -  2\pi r^{a-1} \sum_{k=0}^{-(n+1)} {a\choose k}
   \int^\infty_{r/\eps} 
   \left(-{\eps\rho\over r}\right)^n \left(-{\eps^2\rho^2\over
   r^2}\right)^k \Phi_n(\rho) \rho\, d\rho & if $-a\leq n\leq-1$ \cr
   -2\pi r^{a-1} \int^\infty_{r/\eps} \left(-{\eps\rho\over r}\right)^n
   \left( 1-{\eps^2\rho^2\over r^2} \right)^a \Phi_n(\rho) \rho\,d\rho & if
   $n\leq-(a+1)$ \cr 
}$$

We are interested in estimates for both $r<\eps R_0$ and $r>\eps R_0$;

To estimate $\tilde f$ for $r<\eps R_0$ we begin by estimating the
integrals
$$ \eqalign{
   \alpha_{n,k}(r)&=2\pi \int^{r/\eps}_0 \rho^{k+1} \rho^{k+1} \Phi_n(\rho)
   \,d\rho \qquad k\geq0 \cr
   \beta_{n,k}(r)&=2\pi \int^\infty_{r/\eps} \rho^{k+1} \rho^{k+1}
   \Phi_n(\rho) \,d\rho \qquad k\leq1 \cr
}$$

It is easily seen that
$$ \eqalign{
   \left| \alpha_{n,k}(r) \right| &\leq M {r^{k+2}\over \eps^{k+2}} \qquad 
   k\geq0 \cr 
   \left|\beta_{n,k}(r) \right| &\leq M {r^{k+2}\over \eps^{k+2}} \qquad k 
   \leq  3 \cr
}$$
The estimates for $\beta_{n,k}$ for $k\geq-2$ are more delicate; they must
be obtained by Taylor expanding out about $r/\eps=0$;
$$ \eqalign{
   \beta_{n,-2}(r) &= 2\pi\int^\infty_0 \rho^{-1} \Phi_n(\rho) \,d\rho +
   O(r/\eps) \cr 
   \beta_{n,-1}(r) &= 2\pi\int^\infty_0 \Phi_n(\rho) \,d\rho + O(r/\eps) \cr
   \beta_{n,0}(r)  &= 2\pi\int^\infty_0 \rho \Phi_n(\rho) \,d\rho +
   O(r^2/\eps^2) \cr
   \beta_{n,1}(r) &= 2\pi\int^\infty_0 \rho^2\Phi_n(\rho) \,d\rho +
   O(r^3/\eps^3) \cr 
}$$

We may use the estimates for $\alpha_{n,k}$ and $\beta_{n,k}$ to obtain
estimates for $F_n$

For $a=1$ we have
$$ F_n = \oldcases{
   (-1)^n {\eps^n\over r^n} \left( \alpha_{n,n} - {\eps^2\over r^2}
   \alpha_{n,n+2} \right) & for $n\geq0$ \cr
   -(-1)^n{\eps^n\over r^n} \left( \beta_{n,n}-{\eps^2\over r^2}
   \beta_{n,n+2} \right) & for $n\leq-2$ \cr
   {\eps\over r} \alpha_{-1,1} + {r\over\eps} \beta_{-1,-1} & for $n=-1$ \cr
}$$
giving that
$$ F_n = \oldcases{
   2\pi \int^\infty_0 \!\rho\, \Phi_{-2}(\rho) \,d\rho + O(r^2/\eps^2) &
   for $n=-2$ \cr
   2\pi {r\over\eps} \int^\infty_0 \Phi_{-1}(\rho) \,d\rho  +
   O(r^2/\eps^2) & for $n=-1$ \cr
   -2\pi {r\over\eps} \int^\infty_0 \Phi_{-3}(\rho) \,d\rho +
   O(r^2/\eps^2) & for $n=-3$ \cr
  O(r^2/\eps^2) & for $n\geq0$ or $n\leq -4$
}$$
and for $a=2$ we have
$$ F_n = \oldcases{
   (-1)^n {\eps^n\over r^{n-1}} \left( \alpha_{n,n} - 2 {\eps^2\over r^2}
   \alpha_{n,n+2} + {\eps^4\over r^4} \alpha_{n,n+4} \right) & for $n\geq0$
   \cr
   -(-1)^n{\eps^n\over r^{n-1}} \left( \beta_{n,n}-2{\eps^2\over r^2}
   \beta_{n,n+2} + {\eps^4\over r^4} \beta_{n,n+4} \right) & for $n\leq-3$
   \cr
   \eps \left(2 \alpha_{-1,1} - {\eps^2\over r^2} \alpha_{-1,3} \right)
   +{r^2\over\eps} \beta_{-1,-1} & for $n=-1$ \cr
   {\eps^2 \over r} \alpha_{-1,2} - {r^3\over\eps^2} \beta_{-1,-2} +
   2r\beta_{-1,0} & for $n=-2$ \cr%
}$$
giving that
$$ F_n = \oldcases{
   \eps\left( 2\pi \int^\infty_0 \!\rho^2\, \Phi_{-3}(\rho)\,d\rho +
   O(r^2/\eps^2) \right) & for $n=-3$ \cr
   \eps\left( 2\pi {r\over\eps}\int^\infty_0 \!\rho\, \Phi_0(\rho) \,d\rho
   + O(r^2/\eps^2) \right) & for $n=-2$ \cr
   \eps \left( -2\pi {r\over\eps} \int^\infty_0 \!\rho\, \Phi_{-4}(\rho)
   \,d\rho + O(r^2/\eps^2) \right) & for $n=-4$ \cr
   \eps O(r^2/\eps^2) & for $n\geq-1$ or $n\leq-5$ \cr
}$$

Therefore we have
$$ \eqalign{
   \widetilde{e^{2i\phi}} &= 2\pi \int^\infty_0 \!\rho\, \Phi_{-2}\,d\rho +
   2\pi {re^{i\phi}\over\eps} \int^\infty_0 \Phi_{-1}(\rho)\,d\rho \cr
   & \qquad - 2\pi{re^{-i\phi}\over\eps}
   \int^\infty_0\Phi_{-3}(\rho)\,d\rho + O(r^2/\eps^2) \cr
   \widetilde{re^{3i\phi}} &= 2\pi \eps \int^\infty_0 \!\rho^2\,
   \Phi_{-3}(\rho)\,d\rho + 2\pi re^{i\phi} \int^\infty_0 \!\rho\,
   \Phi_{0}(\rho)\,d\rho \cr
   & \qquad - 2\pi re^{-i\phi} \int^\infty_0 \!\rho\,
   \Phi_{-4}(\rho)\,d\rho + O(r^2/\eps) \cr
}$$

On the other hand if $r>\eps R_0$ then
$$ F_n(r) = \oldcases{
   2\pi r^{a-1} \int^\infty_0 \left(-{\eps\rho\over r}\right)^n
   \left( 1-{\eps^2\rho^2\over r^2} \right)^a \Phi_n(\rho) \rho\,d\rho, &
   if $n\geq0$, \cr
   2\pi r^{a-1} \sum_{k=-n}^a {a\choose k} \int^\infty_0
   \left(-{\eps\rho\over r}\right)^n \left(-{\eps^2\rho^2\over
   r^2}\right)^k \Phi_n(\rho) \rho\, d\rho, & if $-a\leq n \leq-1$, \cr
   0, & otherwise. \cr
}$$
On writing the moment conditions, for $\Phi\in\CA{q}(\Real^2)$, in polar
coordinates as
$$ \eqalign{
   & 2\pi \int^\infty_0 \Phi_0(\rho) \rho \,d\rho =1 \cr
   & 2\pi \int^\infty_0 \rho^{c+1} \Phi_n(\rho) \,d\rho =0 \qquad 0\neq
   |n|\leq c \leq q, \quad \hbox{$c+n$ even};\cr
}$$
it is easily seen that
$$ \eqalign{ F_0(r) &= 1 + O(\eps^{q+1}/r^{q+1}) \cr F_n(r) &=
   O(\eps^{q+1}/r^{q+1}), \qquad n\neq0 \cr }$$
and hence that
$$ \eqalign{
   \widetilde{e^{2i\phi}} &= e^{2i\phi} + O\left(\eps^{q+1}\over
   r^{q+1}\right) \cr
   \widetilde{re^{3i\phi}} &= re^{3i\phi} + O\left(\eps^{q+1}\over
   r^{q}\right) \cr
}$$

\head Acknowledgement \endhead
The authors wish to thank ESI for supporting their visit to the institute
as part of the project on {\it Nonlinear Theory of Generalised functions}.
The authors thank M.~Kunzinger, M.~Oberguggenberger and R.~Steinbauer for
helpful discussions.  J.~Wilson acknowledges the support of
EPSRC. grant No. GR/82236.

\enddocument

\bye